\documentclass[aip,apl,12pt,reprint]{revtex4-1}

\usepackage{graphicx}
\usepackage[mathletters]{ucs}
\usepackage[utf8x]{inputenc}

\begin{document}

\title{Controlled fabrication of single-walled carbon nanotube electrodes by electron-beam-induced oxidation}

\author{Cornelius Thiele}
\email[]{Cornelius.Thiele@kit.edu}

\author{Michael Engel}
\affiliation{Institut f\"ur Nanotechnologie, Karlsruhe Institute of Technology, 76021 Karlsruhe, Germany}
\affiliation{DFG Center for Functional Nanostructures (CFN), 76028 Karlsruhe, Germany}

\author{Frank Hennrich}
\affiliation{Institut f\"ur Nanotechnologie, Karlsruhe Institute of Technology, 76021 Karlsruhe, Germany}

\author{Manfred M. Kappes}
\affiliation{Institut f\"ur Nanotechnologie, Karlsruhe Institute of Technology, 76021 Karlsruhe, Germany}
\affiliation{DFG Center for Functional Nanostructures (CFN), 76028 Karlsruhe, Germany}

\author{Klaus-Peter Johnsen}
\affiliation{Physikalisch-Technische Bundesanstalt, 38116 Braunschweig, Germany}
\author{Carl G. Frase}
\affiliation{Physikalisch-Technische Bundesanstalt, 38116 Braunschweig, Germany}

\author{Hilbert v. L\"ohneysen}
\affiliation{DFG Center for Functional Nanostructures (CFN), 76028 Karlsruhe, Germany}
\affiliation{Physikalisches Institut, Karlsruhe Institute of Technology, 76128 Karlsruhe, Germany}
\affiliation{Institut f\"ur Festk\"orperphysik, Karlsruhe Institute of Technology, 76021 Karlsruhe, Germany}

\author{Ralph Krupke}
\email[]{krupke@kit.edu}
\affiliation{Institut f\"ur Nanotechnologie, Karlsruhe Institute of Technology, 76021 Karlsruhe, Germany}
\affiliation{DFG Center for Functional Nanostructures (CFN), 76028 Karlsruhe, Germany}

\date{\today}

\begin{abstract}

The fabrication of metallic single-walled carbon nanotube electrodes separated by gaps of typically 20\,nm width by electron-beam-induced oxidation is studied within an active device configuration. The tube conductance is measured continuously during the process. The experiment provides a statistical evaluation of gap sizes as well as the electron dose needed for gap formation. Also, the ability to precisely cut many carbon nanotubes in parallel is demonstrated.

\end{abstract}


\maketitle 

\section{Introduction}

In recent experiments, individual carbon nanotubes have been used as true nanoscale electrodes for contacting different types of materials such as nanocrystals\cite{nanogap2004pentacene}, single molecules\cite{nanogap2006o2plasma,breakdown2010cnts}, and phase-change materials\cite{nanogap2011popphasechange}.
The carbon nanotube electrodes were thereby fabricated by making  gaps in pristine nanotubes either using current-induced oxidation or plasma oxidation through masks. A challenge to those  techniques is the limited reliability in fabricating small and reproducible gap sizes. Recently it has been shown that nanogaps in carbon nanotubes can be formed by electron-beam-induced oxidation within a scanning electron microscope\cite{cutting2005ebeamprecision,cutting2006cntsinsidesem,cutting2009ebeamcarbon}. However, this technique has not been used to fabricate single-walled carbon nanotube (SWNT) electrodes within a device geometry. Also the detailed mechanism that leads to gap formation remains vague. In this work, we study the process of gap formation in metallic SWNTs (mSWNTs) within a three-terminal device configuration by electron-beam-induced oxidation (EBIO), while simultaneously measuring the tube conductance. The experiment provides the critical dose for gap formation, a statistical evaluation of the gap size, and identifies the type of electrons that lead to local oxidation.

\section{Experimental Details}

Devices were fabricated on a degenerately doped silicon backgate with 800\,nm of thermal oxide and on 50\,nm thick Si$_3$N$_4$ membranes of 50\,x\,50\,µm$^2$ lateral size, supported by a silicon substrate frame. 200\,nm wide contacts were prepared by standard electron-beam lithography and metallization with 5\,nm titanium and 50\,nm palladium. The gap between metallic contacts was adjusted to the length distribution of the nanotube dispersion and was typically 700\,nm wide.

mSWNTs were prepared by pulsed laser vaporization followed by sonication in 1\,wt\% sodium cholate in D$_2$O and subsequent sorting using density-gradient centrifugation and size-exclusion chromatography as has been reported elsewhere\cite{sorting2002plvtubes, sorting2008hersamreview,sorting2009hennrichsds}. mSWNTs with a diameter distribution of 1.2$\,\pm\,$0.2\,nm and a length of 1\,$\pm$\,0.5\,$\mu$m were used for the nanogap devices.

mSWNTs were deposited onto the metal contacts by dielectrophoresis\cite{cnts2009aravindhighdensity}. For the single-tube devices, a peak-to-peak voltage of V$_{pp}$\,=\,1.5\,V and a frequency of f\,=\,300\,kHz were used for a duration of 3 minutes, while a drop of nanotube dispersion with a concentration of $\sim$5 nanotubes per $\mu$m$^3$ was placed onto the device. For the fabrication of mSWNT thin-film devices, a discontinuous alternating voltage with V$_{pp}$\,=\,6\,V, t$_{on}$\,=\,10\,ms and t$_{off}$\,=\,90\,ms was used to prevent electroosmosis as explained in \cite{cnts2007electroosmosis}.

Imaging and cutting of the nanotubes were carried out in a Zeiss Ultra Plus scanning electron microscope, with the experimental setup shown in Fig. 1a. The built-in charge compensation module was used to inject oxygen gas (purity 99.998\,\%) through a needle of 500\,$\mu$m diameter. The lateral position of the tip of the needle was adjusted at ~200\,$\mu$m off the center of the electron-beam scan window, while its vertical position was fixed to less than 50\,$\mu$m above the device surface.

Gas flow rate into the microscope was monitored by a mass flow controller and set to 35\,$\pm$\,2\,sccm/min at a system pressure of 5.7\,$\cdot$\,10$^{-3}$\,mbar. Molecular flux at the exit of the needle was calculated to 8\,$\cdot$\,10$^7$\,nm$^{-2}$s$^{-1}$. The local pressure at that point was estimated to 1\,mbar. A Keithley 2636A two-channel source measure unit was used in combination with two MM3A-EM Kleindiek nanoprobers mounted inside the electron microscope for in-situ electrical characterization of the nanotube contacts.

Initially, the devices were current-annealed in vacuum to release charges trapped in the substrate from the lithographic fabrication of the contacts\cite{cnts2008metalinsulator}. The current-annealed mSWNT devices showed linear current-voltage (I-V) curves and no gate dependence. I-V curves were measured before and after EBIO gap formation. During the EBIO process a constant source-drain voltage V$_{SD}$ of maximum 1\,V was applied and the current continuously measured in intervals of 50\,ms. 

To cut the carbon nanotubes an electron-beam line scan was executed across the nanotube while injecting the oxygen. During line scans the microscope magnification was adjusted to be either 25\,kX or 50\,kX, yielding a line scan width of 4.57\,$\mu$m and 2.29\,$\mu$m, respectively. The primary electron-beam current used was $\sim$100\,pA, which yields a line dose of $\sim$21.8\,$\mu$C/m and $\sim$43.7\,$\mu$C/m per second, respectively. Acceleration voltage of the primary electrons (PEs) was set to 10\,kV. Scale-calibrated images were used to assess reproducibility of gap sizes.

\begin{figure*}
\includegraphics{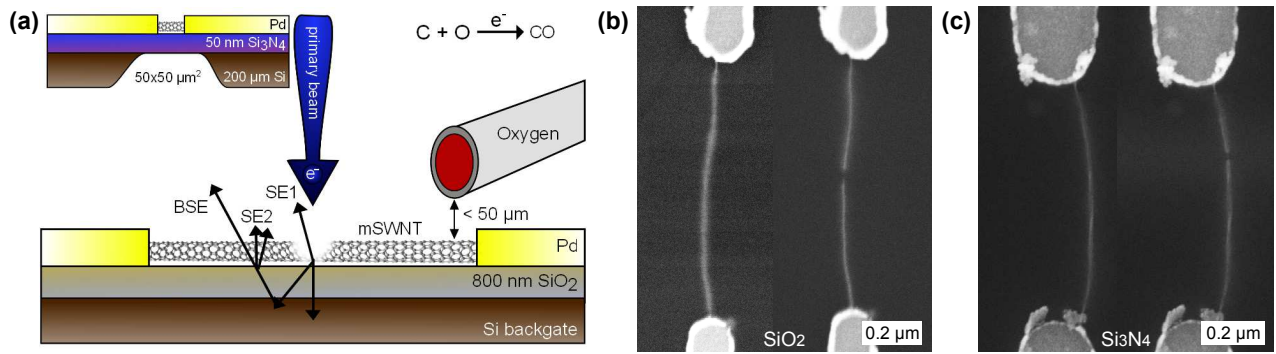}
\caption{\label{figure1} (Color online) (a) Schematic of the EBIO setup: Primary-beam electrons (PE) are repeatedly scanned across a mSWNT in the presence of injected oxygen. The mSWNTs are wired to Pd electrodes and supported by SiO$_2$/Si substrates or Si$_3$N$_4$ membranes (inset). Indicated are the trajectories of secondary (SE1, SE2) and backscattered (BSE) electrons; and the reaction that causes gap formation. 
(b) SE image of a mSWNT on SiO$_2$ before (left) and after (right) gap formation, recorded with V$_g$\,=\,+10\,V. (c) SE image of a mSWNT on a Si$_3$N$_4$ membrane before (left) and after (right) gap formation.}
\end{figure*}

\begin{figure*}
\includegraphics{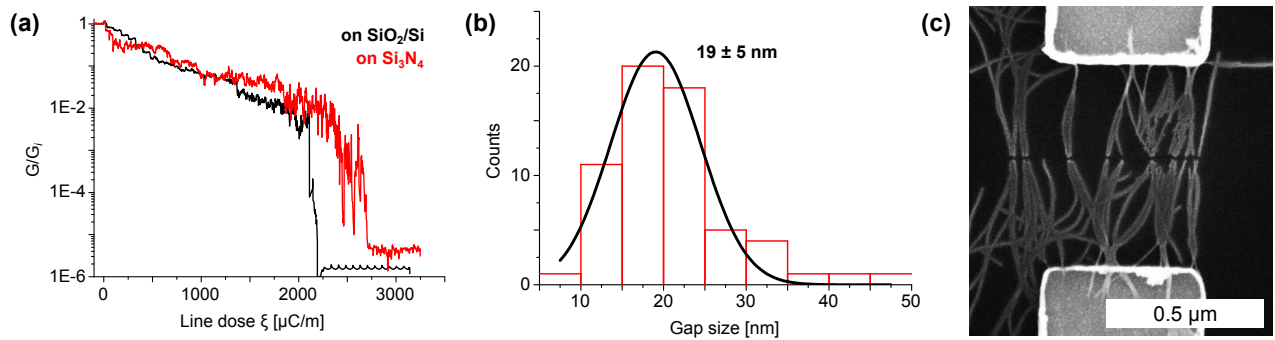}
\caption{\label{figure2} (Color online) (a) Normalized conductance G/G$_i$ vs. line dose $\xi$ recorded during the EBIO process for a device on SiO$_2$/Si and on a Si$_3$N$_4$ membrane. (b) Histogram of nanogap sizes of 62 devices on SiO$_2$/Si substrates, fitted to a normal distribution with µ\,=\,19\,nm and $\sigma$\,=\,5\,nm. (c) SE image of a mSWNT thin-film device after gap formation.}
\end{figure*}

\section{Results and Discussion}

Fig. 1b shows SEM images of an individual mSWNT on
SiO$_2$/Si before (left) and after (right) EBIO-induced gap formation. The images were recorded with an
in-lens detector that is expected to be sensitive to SE1 and SE2 secondary electrons\cite{ebeam1995inlens}. For better visibility of the mSWNT, we used voltage-contrast scanning electron microscopy (VCSEM) to suppress the SEs of the SiO$_2$/Si
substrate\cite{cnts2008aravindvoltagecontrast}. The required backgate voltage V$_g$\,=\,+10\,V was applied with
respect to the source-drain electrodes only for imaging. An effect of V$_g$ on the EBIO process was not observed. 
The EBIO-induced gap shown in Fig. 1b is $\sim$~20\,nm wide. This is a typical result since
the gap size average over 62 devices on SiO$_2$/Si substrates is (19\,$\pm$\,5)\,nm, as shown in Fig. 2b.
Similar results were also obtained with mSWNTs on 50\,nm thick Si$_3$N$_4$ membranes. Fig. 1c shows such
a device on Si$_3$N$_4$ before (left) and after (right) EBIO-induced gap formation. Although recorded
without VCSEM, the membrane appears dark due to its limited SE2 generation\cite{goldstein2003scanning}. We should note that the formation of a gap has also been confirmed by topography measurements with an atomic
force microscope (not shown).
The critical line dose ξ$_C$ that is required for gap formation has been determined from in-situ
conductance measurements. Fig. 2a shows the typical conductance reduction of a mSWNT-device
on SiO$_2$/Si and Si$_3$N$_4$, during the EBIO process. The traces of the conductance G have been normalized to their initial values G$_i$, at the beginning of the EBIO process at t\,=\,0. G$_i$ is typically on the order of (40\,kΩ)$^{-1}$ after current-annealing. G decreases with ξ almost exponentially by two orders of magnitude, before a sudden drop by additional 3-4 orders of magnitude down to the electron-beam-induced residual
conductance is recorded $<$(10\,GΩ)$^{-1}$. This last step indicates the formation of a gap in the nanotube.
There are no plateaus in G which could indicate a step-wise introduction of defects. Typically ξ$_C$ is on
the order of a few mC/m. This behavior is in marked contrast to the continuous electron-beam-induced metal-to-insulator transitions, which are caused by charges trapped in the substrate surface\cite{cnts2008metalinsulator}. It
has been shown that such transitions require a line dose of $\sim$200\,µC/m at 10\,kV PE energy. During the EBIO process, such a small dose causes only a small reduction in G (see Fig. 2a), and cutting of tubes
requires a dose which is ten times as high. We assume that charging is not important here since
oxygen ions can compensate electron-beam-induced surface charges. Note also that it has been
shown that trapped charges can be released under large bias and that the original conductance can
be fully restored (reversibility)\cite{cnts2008metalinsulator}. In the present case - due to the gap formation - the conductance drop remains irreversible even at V$_{SD}$\,$\gg$\,10\,V.

We now want to discuss the mechanism that leads to gap formation. Direct knock-on damage can be
excluded, as the threshold for carbon nanotubes is 85\,keV\cite{cnts1999knockondmg}, much higher than the electron
energies used in this and in previous work\cite{cutting2005ebeamprecision}. We have also tested the use of argon instead of oxygen, but could not open a gap. It is likely that oxygen radicals forming under the electron beam
are causing the gap formation. The lack of a significant difference between EBIO on a thick SiO$_2$/Si
substrate and a thin Si$_3$N$_4$ membrane implies that backscattered electrons (BSE) and BSE-generated
secondary electrons (SE2) play a minor role since Monte Carlo simulations \cite{spieproceedings2010frase} showed that almost no electrons are backscattered from the Si$_3$N$_4$ membrane and its total SE yield (SE1 + SE2) is strongly reduced due to the absence of BSE and thus SE2. Moreover, the EBIO process appears to be insensitive to a
suppression of SE1 and SE2 electrons by VCSEM. This indicates that the PEs are causing the gap formation. 

A recent work on electron-beam-induced etching of SiO$_2$ using XeF$_2$ gas\cite{cutting2008continuummodel} indicates that the source of oxygen radicals is effectively limited to the area where the PE beam hits the surface, which for a line scan would be a strip of $\sim$1-2\,nm width. What causes the gap to become $\sim$20\,nm wide remains unclear at the moment. We assume that the gap acquires this large width either because of an exothermic chain reaction where the gap size is determined by the oxygen partial pressure in the system\cite{breakdown2010cnts}, or due to the diffusion of oxygen radicals. In both cases a variation of the pressure should have an influence on the gap size, which will be investigated in the future.

As an interesting final observation, we note EBIO is not limited to the cutting of individual nanotubes, but can also be extended to form gaps in devices with multiple nanotubes. Fig. 2c shows such an example. The device which
contains $\sim$50 mSWNTs was produced by pulsed dielectrophoresis. With EBIO, gaps of similar size
can be produced in all of the nanotubes at the same relative position. Such a result cannot be
obtained by current-induced oxidation\cite{breakdown2011ultrahighdensity, 2011khondakerAPL, breakdown2011hotspots_pop}. 

\section{Conclusions}

Reliable and lithographically precise fabrication of SWNT electrodes by electron-beam-induced oxidation was demonstrated for individual SWNT and SWNT thin-film devices. Nanogaps can be fabricated reliably down to a size of 19\,$\pm$\,5\,nm. The independence of the gap size of the choice of substrate and of the number of secondary electrons suggests that the process is driven by primary electrons. Although the gaps produced by EBIO are still large compared to molecular scales, they have an advantageous size for addressing phase-change materials\cite{nanogap2011popphasechange} where reproducible gap sizes are required to
obtain reproducible switching properties.

\end{document}